\documentclass[bibnotes,pra,twocolumn,superscriptaddress]{revtex4}%
\usepackage{epsfig,dsfont,amssymb,amsmath,amsthm,amsfonts,amsbsy,mathrsfs}
\usepackage{graphicx}
\usepackage{bm}

\begin{document}
\title{Dynamical Phonon Laser in Coupled Active-Passive Microresonators}
\author{Bing He}
\affiliation{Department of Physics, University of Arkansas, Fayetteville, Arkansas 72701, USA}
\author{Liu Yang }
\affiliation{Harbin Engineering University, College of Automation, Harbin, Heilongjiang 150001, China
}
\author{Min Xiao}
\affiliation{Department of Physics, University of Arkansas, Fayetteville, Arkansas 72701, USA}
\affiliation{National Laboratory of Solid State Microstructures and School of Physics, Nanjing University, Nanjing 210093, China}

\begin{abstract}
Effective transition between the population-inverted optical eigenmodes of two coupled microcavities carrying mechanical oscillation realizes 
a phonon analogue of optical two-level laser. By providing an approach that linearizes the dynamical equations of weak nonlinear systems without relying on their steady states, we study such phonon laser action as a realistic dynamical process, which exhibits time-dependent stimulated phonon field amplification especially when one of the cavities is added with optical gain medium. The approach we present explicitly gives the conditions 
for the optimum phonon lasing, and thermal noise is found to be capable of facilitating the phonon laser action significantly. 
\end{abstract}
\maketitle

\begin{figure}[b!]
\vspace{0cm}
\centering
\epsfig{file=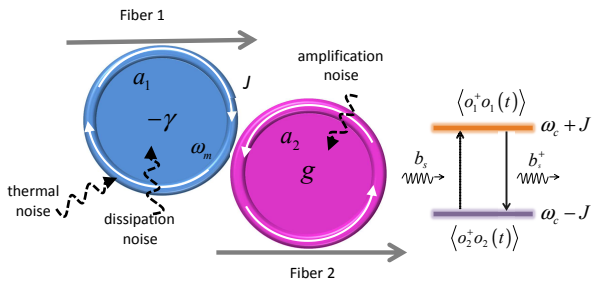,width=0.9\linewidth,clip=} 
{\vspace{-0.2cm}\caption{(color online) Setup of coupled microcavities with their coupling rate $J$ adjusted by their gap distance. The first cavity carries a mechanical mode. The pump field from the second optical fiber for amplification does not couple to the first cavity. The stimulated transition of phonons takes place between two supermode states $\hat{o}^\dagger_1|0\rangle$ and $\hat{o}^\dagger_2|0\rangle$ separated by an energy level difference $2J$, and their occupation numbers $\langle \hat{o}^\dagger_i\hat{o}_i(t)\rangle$ ($i=1,2$) are generally time-dependent in the dynamical operation of the setup. }}
\vspace{0cm}
\end{figure}

Compound structures like coupled microcavities or waveguides constitute large number of interesting systems in optical sciences. An important category 
that has recently attracted extensive researches covers those with alternately distributed active (gain) and passive (loss) components, as they can 
mimic the parity-time ($\mathcal {PT}$) symmetric quantum mechanics \cite{bender}, a generalization of ordinary quantum mechanics. In addition to the theoretical investigations (see, e.g. \cite{w1,k, sc, ch, oth,arga, ben2, l-l, m1,m2, m3, bhe0}), numerous experiments have demonstrated peculiar features of light transmission in these systems \cite{ex1, ex2,ex3, ex4, ex5, ex6, ex7}. Richer phenomena could manifest if they incorporate other degrees of freedom to form hybrid systems, which have been studied by combining $\mathcal {PT}$ symmetric systems with Kerr nonlinearity \cite{n1,n2,n3,n5,n7,bhe} and mechanical oscillators \cite{jing,oms1,oms2, oms3}. 

The device of two coupled microcavities in Fig. 1 can implement phonon laser action \cite{l-4}, as well as in many other systems \cite{l-1,l-2,l-21, l-3,l-5, l-6, l-7,l-71, l-8, l-81, l-9, l-10}. 
Here the coupling intensity $J$ of the two microcavities is determined with their adjustable gap. Under a pump drive of the intensity $E$ and frequency $\omega_L$, two eigenmodes or supermodes of different energy levels, as the superpositions of the individual cavity modes, will be built up. If one of the cavities also carries a mechanical oscillation with the frequency $\omega_m$, the cavity supermodes will couple to the associated phonon field in cavity material via radiation pressure. Once there is a population inversion between the cavity supermodes, an amplification of the phonon field will be realized in analog to an optical laser. 

A recent study \cite{jing} proposes the enhancement of the phonon lasing by adding optical gain medium into one of the cavities (also see \cite{oms3} 
for a continued study in the similar approach). Then the system will have the exact $\mathcal {PT}$ symmetry given the equal gain rate $g$ and loss rate $\gamma$ of the respective cavities, and this $\mathcal {PT}$ symmetric point was predicted to be capable of achieving the best performance of the phonon laser driven by resonant pump \cite{jing}. A prediction like this was made under the assumption that the phonon laser operates in a steady state, in which the expectation values of the cavity modes $\hat{a}_1, \hat{a}_2$ and mechanical mode $\hat{b}$ keep unchanged with time. 

However, as we will show below, the phonon laser should operate under a blue detuned pump which leads to no steady state. In the presence of optical gain the similar systems can be fully dynamical. A well-known example is that, at the above mentioned $\mathcal {PT}$ symmetric point $g=\gamma$, the intracavity light fields are totally variable, exhibiting a transition from periodically oscillating to exponentially growing as the cavity coupling $J$ decreases across the exceptional point $J=\gamma$. A slight change of a cavity's size under radiation pressure can hardly make these dynamically evolving fields become time-independent. Properly understanding the concerned phonon laser operation necessitates an approach based on dynamical picture.

To be more specific, the system's dynamical equations read \cite{note}  
\begin{eqnarray}
&&\dot{\hat{a}}_1=-(\gamma-ig_m\hat{x})\hat{a}_1-iJ\hat{a}_2 +E e^{-i\Delta t}+\sqrt{2\gamma}\hat{\xi}_p,\\
&&\dot{\hat{a}}_2=g \hat{a}_2-iJ\hat{a}_1 +\sqrt{2 g}\hat{\xi}^\dagger_a,\\
&&\dot{\hat{b}}=-\gamma_m \hat{b}-i\omega_m \hat{b}+ig_m\hat{a}_1^\dagger\hat{a}_1+\sqrt{2\gamma_m} \hat{\xi}_m
\label{equation}
\end{eqnarray}
in a frame co-moving at the frequency $\omega_c$ ($\Delta=\omega_c-\omega_L$) of the two cavities, where $\hat{x}=\hat{b}+\hat{b}^\dagger$ is the dimensionless position operator of the mechanical oscillator damping at the rate $\gamma_m$ and coupled to the passive mode occupation $\hat{a}_1^\dagger\hat{a}_1$ with a constant $g_m=\omega_c x_0/R$ ($x_0$ is the the mechanical oscillator's zero point fluctuation and $R$ is the cavity size). Without a classical steady state it will be impossible to linearize the dynamical equations (1)-(3) following the practice in most other works about quantum optomechanics. Moreover, these equations carry the random drive terms of the dissipation (amplification) noise $\hat{\xi}_p$ ($\hat{\xi}_a$) and the thermal noise $\hat{\xi}_m$, which satisfy the relations $\langle \hat{\xi}_i(t) \hat{\xi}_i^\dagger (t')\rangle=\delta(t-t')$ ($i=p,a$) and $\langle \hat{\xi}_m(t) \hat{\xi}_m^\dagger (t')\rangle=(n_{th}+1)\delta(t-t')$ ($n_{th}$ is the thermal reservoir mean occupation number). The effects of these quantum noises, which are neglected in the previous studies but exist in any concerned quantum dynamical process, should be well clarified. In this work we develop an approach to such quantum dynamical processes. The population inversion of the optical supermodes, as the key to the phonon lasing, will be determined in this approach capable of dealing with the quantum noises which are indispensable as shown below.

Our approach makes use of the stochastic Hamiltonian 
\begin{eqnarray}
H_{SR}(t)&=&i\big\{\sqrt{2\gamma }(\hat{a}_{1}^{\dagger }\hat{\xi}
_{p}(t)-H.c.)
+\sqrt{2g }(\hat{a}_{2}^{\dagger }\hat{\xi}_{a}^{\dagger }(t)-H.c.)\nonumber\\
&+&\sqrt{2\gamma _{m}}(\hat{b}^{\dagger }\hat{\xi}_{m}(t)-H.c.)\big\} 
\end{eqnarray}
in terms of the system-reservoir couplings for the amplification and dissipations in the system (the notation in \cite{bhe} for the amplification part is adopted). The quantum dynamical equations (1)-(3) can be obtained by the small increments $d\hat{o}(t)=U^\dagger(t+dt,t)\hat{o}(t)U(t+dt,t)-\hat{o}(t)$ of the operators $\hat{o}=\hat{a}_1,\hat{a}_2$ and $\hat{b}$, which are under 
the evolution $U(t)={\cal T}\exp\{-i\int_0^t d\tau [H_S(\tau)+H_{OM}+H_{SR}(\tau)]\}$ of the total Hamiltonian \cite{book}. The Hamiltonians inside the time-ordered exponential include the part 
\begin{eqnarray}
H_S(t)&=&\omega _{c}\hat{a}_{1}^{\dagger }\hat{a}_{1}+\omega _{c}\hat{a}_{2}^{\dagger }\hat{a}_{2}+\omega _{m}\hat{b}^{\dagger }\hat{b}+J(\hat{a}_{1}\hat{a}_{2}^{\dagger }+\hat{a}_{1}^{\dagger }\hat{a}_{2})\nonumber\\
&+&iE(\hat{a}_{1}^{\dagger }e^{-i\omega _{L}t}-\hat{a}_{1}e^{i\omega _{L}t})
\end{eqnarray}
about the cavity coupling plus the external drive, as well as the one $H_{OM}=-g_m\hat{a}_1^\dagger\hat{a}_1(\hat{b}+\hat{b}^\dagger)$ about optomechanical interaction.  

We apply an interaction picture with respect to the system Hamiltonian $H_S(t)$, whose action $U_0(t)=\mathcal{T}\exp\{-i\int_0^t d\tau H_S(\tau)\}$ evolves the cavity modes as the exact transformation
\begin{eqnarray}
\left( 
\begin{array}{c}
U^\dagger_0\hat{a}_1 U_0\\ 
U^\dagger_0\hat{a}_2U_0
\end{array}
\right) 
&=&\frac{1}{\sqrt{2}}e^{-i\omega_ct}\left( 
\begin{array}{c}
\frac{\hat{a}_1+\hat{a}_2}{\sqrt{2}}e^{-iJt}+\frac{\hat{a}_1-\hat{a}
_2}{\sqrt{2}}e^{iJt} \\ 
\frac{\hat{a}_1+\hat{a}_2}{\sqrt{2}}e^{-iJt}-\frac{\hat{a}_1-\hat{a}_2}{\sqrt{2}}e^{iJt}%
\end{array}
\right)\nonumber\\
 &+&
\sqrt{2}e^{-i\omega_ct}\left( \begin{array}{c}
E_1(t) \\ 
E_2(t)
\end{array}
\right),  \label{l-solution}
\end{eqnarray}
where 
\begin{eqnarray}
E_1(t)&= &\frac{iE}{2\sqrt{2}}(\frac{1}{\Delta+J}e^{-iJt} +\frac{1}{\Delta-J}e^{iJt}\nonumber\\
& -&\frac{2\Delta}{\Delta^2-J^2}e^{i\Delta t}),\nonumber\\
E_2(t)&=&\frac{iE}{\sqrt{2}}\big(\frac{J}{\Delta^2-J^2}e^{i\Delta t}-\frac{J}{\Delta^2-J^2}\cos(Jt)\nonumber\\
&-&i\frac{\Delta}{\Delta^2-J^2}\sin(Jt)\big).
\label{E}
\end{eqnarray}
The optical supermodes $\hat{o}_{1,2}=(\hat{a}_1\pm \hat{a}_2)/\sqrt{2}$ with the energy levels $\omega_c\pm J$ naturally appear in Eq. (\ref{l-solution}). Taking the interaction picture is equivalent to the factorization
\begin{eqnarray}
&&{\cal T}e^{-i\int_0^t d\tau (H_S(\tau)+H_{OM}+H_{SR}(\tau))}\nonumber\\
&=&U_0(t)~{\cal T}e^{-i\int_0^t d\tau U_0^\dagger(\tau)(H_{OM}+H_{SR}(\tau))U_0(\tau)}
\end{eqnarray}
of the evolution operator $U(t)$ \cite{fac}, to have the exact form $U_0^{\dagger}(t)(H_{OM}+H_{SR}(t))U_0(t)$ 
in one of the time-ordered exponentials above consisting of two parts. 
One is in a time-dependent quadratic form plus a mechanical 
displacement term and three system-reservoir coupling terms
\begin{eqnarray}
H_1(t)&=&-g_m \{[E_1(t)(\hat{o}^\dagger_1 e^{iJt}+\hat{o}
^\dagger_2 e^{-iJt})+H.c.]\nonumber\\
&+&2|E_1(t)|^2\}(\hat{b}e^{-i\omega_m t}+\hat{b}^\dagger e^{i\omega_m t})\nonumber\\
&+& i\sqrt{g}\{(\hat{o}^\dagger_1 e^{iJt}-\hat{o}^\dagger_2
e^{-iJt}+2E_2^\ast(t))e^{i\omega_c t}\hat{\xi}^\dagger_a-H.c.\}\nonumber\\
&+& i\sqrt{\gamma}\{(\hat{o}^\dagger_1 e^{iJt}+\hat{o}^\dagger_2
e^{-iJt}+2E_1^\ast(t))e^{i\omega_c t}\hat{\xi}_p-H.c.\}\nonumber\\
&+& i\sqrt{2\gamma_m}\{\hat{b}^\dagger e^{i\omega_m t}\hat{\xi}_m(t)-H.c.\},
\label{1}
\end{eqnarray}
and the other is the cubic nonlinear one
\begin{eqnarray}
H_2(t)&=&-\frac{1}{2}g_m(\hat{o}^\dagger_1 e^{iJt}+\hat{o}^\dagger_2 e^{-iJt})(\hat{o}_1 e^{-iJt}+\hat{o}_2
e^{iJt})\nonumber\\
&\times & (\hat{b}e^{-i\omega_m t}+\hat{b}^\dagger e^{i\omega_m t}).  
\label{2}
\end{eqnarray}
The terms containing $\hat{o}_1\hat{o}_2^\dagger b^\dagger$ or its conjugate in the second Hamiltonian $H_2(t)$ indicate a transition from the blue supermode $\hat{o}_1$ to the red supermode $\hat{o}_2$ while generating a phonon (see the level scheme in Fig. 1), realizing phonon lasing once the occupation of the blue supermode surpasses that of the red one. The Hamiltonian $H_2(t)$ also gives the resonant transition between the two supermodes at $\omega_m=2J$, i.e. the coefficient of $\hat{o}_1\hat{o}_2^\dagger b^\dagger$ becomes unity, corresponding to the gain spectrum line center of stimulated phonon field \cite{l-4}.

Under the simultaneous action of $H_1(t)$ and $H_2(t)$, the supermode populations 
\begin{eqnarray}
&&\langle \hat{o}_i^\dagger\hat{o}_i(t)\rangle=\mbox{Tr}_S(\hat{o}_i^\dagger\hat{o}_i\rho_S(t))\nonumber\\
&&~~~~~~~~~~~=\mbox{Tr}_S\big\{\hat{o}_i^\dagger\hat{o}_i\mbox{Tr}_R\big(U(t)\rho_S(0)\rho_R U^{\dagger}(t)\big)\big\},~~
\end{eqnarray}
for $i=1,2$, are predominantly determined by the former. Here $\rho_S(t)$ and $\rho_R$ are the reduced system state and the total reservoir state, 
respectively. This can be seen from their following reduction
\begin{eqnarray}
&&\langle \hat{o}_i^\dagger\hat{o}_i(t)\rangle
=\mbox{Tr}_{S,R}\big\{\hat{o}_i^\dagger\hat{o}_i U_0(t)~{\cal T}e^{-i\int_0^t d\tau(H_{1}+H_{2})(\tau)}\nonumber\\
&&~~~~~~~~~~~\times \rho_S(0)\rho_R ~{\cal T}e^{i\int_0^t d\tau(H_{1}+H_{2})(\tau)}U_0^\dagger (t)\big\}\nonumber\\
&\approx&\mbox{Tr}_{S,R}\big\{ U_1^\dagger(t)U_0^\dagger (t)\hat{o}_i^\dagger\hat{o}_i U_0(t)U_1(t)
U_2(t)\rho_S(0)\rho_R U^\dagger_2(t)\big\}\nonumber\\
&=&\mbox{Tr}_{S,R}\big\{ U_1^\dagger(t)U_0^\dagger (t)\hat{o}_i^\dagger\hat{o}_i U_0(t)U_1(t)\rho_S(0)\rho_R\big\},
\label{reduction}
\end{eqnarray} 
where $U_l(t)={\cal T}e^{-i\int_0^t d\tau H_{l}(\tau)}$ for $l=1,2$. In Eq. (\ref{reduction}), the relation $U_2(t)\rho_S(0)U_2^\dagger(t)=\rho_S(0)$ for the system's initial state $\rho_S(0)$, the product of a cavity vacuum state $|0\rangle_c$ and a mechanical thermal state, is due to the fact $H_2(t)|0\rangle_c=0$. The approximate equality in Eq. (\ref{reduction}) comes from factorizing the actions of the non-commutative Hamiltonians $H_1(t)$ and $H_2(t)$ as
\begin{eqnarray}
{\cal T}e^{-i\int_0^t d\tau(H_{1}+H_{2})(\tau)}&=&{\cal T}e^{-i\int_0^t d\tau U_2(t,\tau)H_{1}(\tau)U^\dagger_2(t,\tau)}U_2(t)\nonumber\\
&\approx & U_1(t)U_2(t).
\label{app}
\end{eqnarray}
For the experimentally realizable optomechanical systems of weak coupling, the corrections to the 
system operators by the unitary operation $U_2(t,\tau)={\cal T}e^{-i\int_\tau^t dt' H_2(t')}$ are in the higher orders of the coefficient $g_m/\omega_m\ll 1$ (see the Supplementary Materials), so that they can be well neglected to use the original form of $H_1(\tau)$ in the time-ordered exponential on the right side of the above equation. This only approximation we use in the calculations of the optical supermode populations is independent of the drive intensity $E$. 

While the unitary operation $U_0(t)$ only displaces the supermode operators in Eq. (\ref{reduction}), the action $U_1(t)$ of the Hamiltonian $H_1(t)$ leads to the following dynamical equations \cite{book} 
\begin{eqnarray}
\dot{\hat{o}}_1&=& 1/2(g-\gamma)\hat{o}_1+1/2(g+\gamma) e^{2iJt}\hat{o}_2\nonumber\\
&+& ig_m E_1(t)e^{iJt}(\hat{b}%
e^{-i\omega_m t}+\hat{b}^\dagger e^{i\omega_m t})\nonumber\\
&+&(\gamma E_1(t)-g E_2(t))e^{iJt}+\hat{n}_1(t),\nonumber\\
\dot{\hat{o}}_2&=& 1/2(g+\gamma)e^{-2iJt}\hat{o}_1+1/2(g-\gamma)\hat{o}_2\nonumber\\
&+&ig_m E_1(t)e^{-iJt}(\hat{b}e^{-i\omega_m t}+\hat{b}^\dagger e^{i\omega_m t})\nonumber\\\
&+&(\gamma E_1(t)+g E_2(t))e^{-iJt}+\hat{n}_2(t),\nonumber\\
\dot{\hat{b}}&=& -\gamma_m \hat{b}+ig_m E^\ast_1(t)e^{i\omega_m t}(\hat{o}_1 e^{-iJt}+\hat{o}_2 e^{iJt})\nonumber\\
&+& ig_m E_1(t)e^{i\omega_m t}(\hat{o}%
^\dagger_1 e^{iJt}+\hat{o}^\dagger_2 e^{-iJt})\nonumber\\
&+&2i g_m|E_1(t)|^2e^{i\omega_m t}+\hat{n}_3(t)
\label{dynamic}
\end{eqnarray}
for the system operators, where 
\begin{eqnarray}
\hat{n}_1(t)&=&\sqrt{g}
e^{iJt}e^{i\omega_c t}\hat{\xi}^\dagger_a(t)+\sqrt{\gamma}e^{iJt}e^{i\omega_c t}\hat{\xi}_p(t),\nonumber\\
\hat{n}_2(t)&=&\sqrt{g} e^{-iJt}e^{i\omega_c t}\hat{\xi}^\dagger_a(t)-\sqrt{\gamma}e^{-iJt}e^{i\omega_c t}\hat{\xi}_p(t),\nonumber\\
\hat{n}_3(t)&=&\sqrt{2\gamma_m}e^{i\omega_m t}\hat{\xi}_m(t).
\label{noise}
\end{eqnarray}
The noise drive terms in Eq. (\ref{noise}) must be included in these equations. For example, in the trivial situation of turning off the pump drive ($E=0$),
the damping of the mechanical mode would result in its ``cooling" to the ground state, i.e. $\langle \hat{b}^\dagger\hat{b}(t)\rangle\rightarrow 0$ as $t\rightarrow \infty$, were there no thermal noise term $\hat{n}_3(t)$ in the last equation of (\ref{dynamic}). The invariant occupation number $\langle \hat{b}^\dagger \hat{b}\rangle$ under such thermal equilibrium is preserved with the complete form $\hat{b}(t)=e^{-\gamma_mt}\hat{b}+\sqrt{2\gamma_m}\int_0^t d\tau e^{-\gamma_m(t-\tau)}e^{i\omega_m\tau}\hat{\xi}_m(\tau)$ of the evolved mechanical mode. The evolved supermodes $\hat{o}_1(t), \hat{o}_2(t)$, on the same footing with $\hat{b}(t)$ in Eq. (\ref{dynamic}), should include the contributions from the quantum noises as well. 

The next question is how to evolve the supermodes so that a good population inversion $\Delta N(t)=\langle \hat{o}_1^\dagger\hat{o}_1(t)\rangle-\langle \hat{o}_2^\dagger\hat{o}_2(t)\rangle$ can be achieved. One advantage of our approach is that the conditions for realizing the optimal population inversion can be straightforwardly read from Eq. (\ref{dynamic}), which is an inhomogeneous system of differential equations with the coherent and noise drive terms. The coefficients of $\hat{o}_i$ or $\hat{o}_i^\dagger$ on the right side of the last equation, for examples, are generally the sums of complex exponential functions of $t$ considering the form of $E_1(t)$. These coefficients reflect the intensities of the beam splitter (BS) type coupling in the form $f(t)\hat{o}_i\hat{b}^\dagger+H.c.$ or the squeezing (SQ) type coupling in the form $g(t)\hat{o}^\dagger_i\hat{b}^\dagger+H.c.$, where the exact functions $f(t),g(t)$ can be found from Eq. (\ref{1}). These couplings can be enhanced if a complex exponential function of $t$ in $f(t)$ or $g(t)$ becomes unity. 
A significant population inversion will be realized if an SQ coupling between the blue supermode $\hat{o}_1$ and the mechanical mode $\hat{b}$ can be strengthened. Such enhancement will be possible by setting the pump to blue sideband with its detuning $\Delta$ equal to $-\omega_m-J=-3J$ [considering the optimal transition condition $\omega_m=2J$ from Eq. (\ref{2})], reducing the factor $e^{i(\Delta+\omega_m+J)t}$ before $\hat{o}_1^\dagger$ in the last equation of (\ref{dynamic}) to a unity. 

\begin{figure}[h!]
\vspace{0cm}
\centering
\epsfig{file=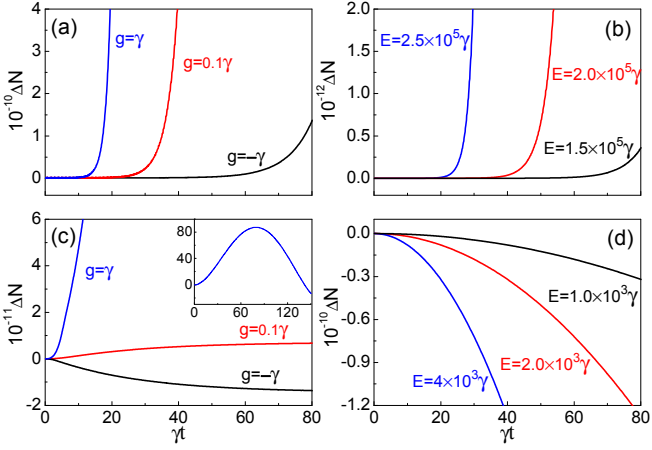,width=1\linewidth,clip=} 
{\vspace{-0.2cm}\caption{(color online) Population inversion evolutions in a setup with the system parameters $\omega_m=22.8\gamma$, $g_m=5\times 10^{-5}\gamma$ and $\gamma_m=0.037\gamma$, which are converted from those of the experiment in \cite{l-4}. The setup operates under the optimal transition condition $\omega_m=2J$ and in an environment of $T=273$K ($n_{th}=2.4\times 10^5$). The choice of the other parameters: (a) $E=2.5\times 10^5\gamma$, $\Delta=-3J$; (b) $g=0.5\gamma$, $\Delta=-3J$; (c) $E=10^7\gamma$, $\Delta=0$; (d) $g=0.5\gamma$, $\Delta=J$. The inserted plot in (c) shows the long-term behavior for the curve of $g=\gamma$.}}
\vspace{-0cm}
\end{figure}

To illustrate the general theory, we plot the population inversions in terms of the dimensionless parameters in Fig. 2. 
These inversions are numerically calculated with Eq. (\ref{dynamic}). Figures 2(a) and 2(b) show that, under the above mentioned two optimal conditions, the inversions grow with time due to the SQ process. Increased gain rate $g$ and drive intensity $E$ serve as the additional factors to make them go up monotonously further. The inversion in a passive setup ($g=-\gamma$) can increase with time, in addition to reaching the steady states (not shown here) under lower drive intensity $E$ for this passive setup (in the absence of considerably high optical gain, steady states may exist under the condition $g_m|\alpha_i|\ll \gamma$ for a blue detuned drive, where $\alpha_i$ are the average cavity field amplitudes proportional to the drive intensity $E$; see e.g. a proposed setup in \cite{laser1}). The enhanced SQ process heats up the cavity material with increased thermal occupation $\langle \hat{b}^\dagger\hat{b}(t)\rangle$ different from the quantity $|\langle \hat{b}(t)\rangle|^2$, and the very strong light fields after a long period will make the system go beyond the current model of linear amplification and dissipation in accordance with the specific material properties.

\begin{figure}[b!]
\vspace{-0cm}
\centering
\epsfig{file=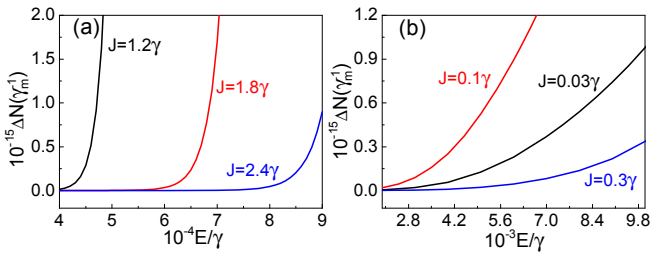,width=1\linewidth,clip=} 
{\vspace{-0.2cm}\caption{(color online) Relations between the realized population inversion at $t=\gamma_m^{-1}$ and the drive intensity, under the conditions $\omega_m=2J$ and $\Delta=-\omega_m-J$. Here we set $g=0.5\gamma$. The regimes of $J>\gamma$ and $J<\gamma$ are illustrated in (a) and (b), respectively. The other fixed system parameters are the same as those in Fig. 2. At a fixed $E$, the inversion increases with the lowered $J$ from $0.3\gamma$ to $0.1\gamma$, but drops as $J$ is decreased further to $0.03\gamma$.}}
\vspace{0cm}
\end{figure}

As comparison we also present two other examples. The first one in Fig. 2(c) is to drive the passive cavity resonantly at $\Delta=0$, having $E_1(t)=iE/(2\sqrt{2}J) (e^{-iJt}-e^{iJt})$. The term with the factor $e^{-iJt}$ in the $E_1(t)$ provides enhanced SQ coupling between $\hat{o}_2$ and $\hat{b}$, while the one with $e^{iJt}$ enhances the BS coupling between $\hat{o}_1$ and $\hat{b}$, showing that the SQ effect will dominate in the end. The other example in Fig. 2(d) has $\Delta=J$, which happens to be one of the resonant points of the coupled system so that $E_1(t)=iE/(4\sqrt{2}J)(e^{-iJt}-2(1+iJt)e^{iJt})$. In this special situation, the extra linearly increasing factor overshadows the effects of the phase factors ($e^{\pm iJt}$) and nonetheless enhances the SQ coupling between $\hat{o}_2$ and $\hat{b}$, to give negative population inversions.

The influence of the cavity coupling intensity on supermode population inversion is illustrated in Fig. 3, showing the relations between the inversion (at the mechanical oscillation lifetime $\gamma^{-1}_m$) and the drive intensity for different $J$. To keep the optimal conditions in Fig. 2(a), the mechanical frequency $\omega_m$ is also assumed to be adjustable in the illustrations. One sees that, given a fixed drive intensity $E$, a lowed coupling $J$ actually increases the population inversion until it becomes small enough to have the two cavities almost decoupled. This is totally contrasting to the prediction of no phonon lasing in the regime $J<(g+\gamma)/2$ by a previous study \cite{jing}. That conclusion is based on the diagonalized form $(\omega_{+}+i\gamma_+)\hat{q}_2^\dagger\hat{q}_2+(\omega_{-}+i\gamma_-)\hat{q}_1^\dagger\hat{q}_1$ of the non-Hermitian Hamiltonian $(\omega_c-i\gamma) \hat{a}_1^\dagger\hat{a}_1+(\omega_c+ig) \hat{a}_2^\dagger\hat{a}_2+J(\hat{a}_1^\dagger \hat{a}_2+\hat{a}_2^\dagger\hat{a}_1)$ widely used in the study 
of $\mathcal{PT}$ symmetric optical systems, suggesting that phonons induce 
a transition between the modes $\hat{q}_1, \hat{q}_2$ with their gap $\omega_+-\omega_-$ disappearing when $J<(g+\gamma)/2$. In fact, these generally non-orthogonal modes (see more detailed discussion in \cite{ex2}) coincide with the supermodes $\hat{o}_1, \hat{o}_2$ only in a special situation of $g=-\gamma$; see the Supplementary Materials. Similar to the transitions between atomic levels, the action of the Hermitian Hamiltonian $H_2(t)$ can only cause an effective transition between two orthogonal states like $\hat{o}^\dagger_1|0\rangle$ and $\hat{o}^\dagger_2|0\rangle$, and the transition between the non-orthogonal states $\hat{q}^\dagger_1|0\rangle$ and $\hat{q}^\dagger_2|0\rangle$ with $\langle 0|\hat{q}_1\hat{q}_2^\dagger|0\rangle\neq 0$ is forbidden for arbitrary system parameters.  

\begin{figure}[h!]
\vspace{-0cm}
\centering
\epsfig{file=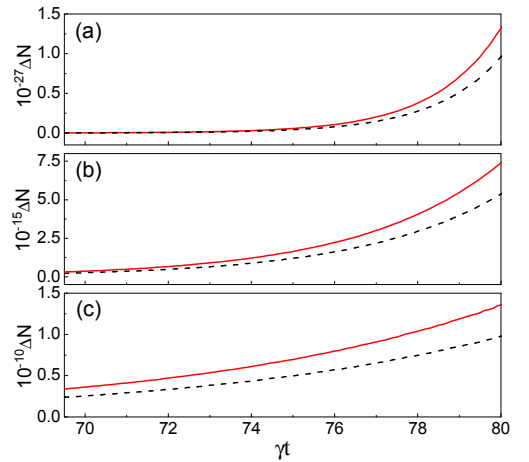,width=0.77\linewidth,clip=} 
{\vspace{-0.2cm}\caption{(color online) Thermal noise contribution to the supermode population inversion. The solid curves in (a), (b) and (c) are the portions of those in Fig. 2(a), with $g=\gamma$, $0.1\gamma$ and $-\gamma$, respectively. The dashed curves represent the contributions from the thermal noise drive $\hat{n}_3$ in Eq. (\ref{noise}).  }}
\vspace{-0cm}
\end{figure}

A unique property of the optical medium is that the quantum noises, which must be considered as mentioned before, can significantly affect the supermode populations. We illustrate this important fact in Fig. 4 showing the proportions of the thermal noise contribution in the results of Fig. 2(a). The detailed calculation of the noise contributions can be found in the Supplementary Materials. It is seen from the comparisons in Fig. 4 that, under the enhanced SQ coupling due to the properly chosen system parameters, the thermal noise acting as a random drive can predominantly contribute to the population inversions. 
The contribution is proportional to the thermal occupation number $n_{th}$, a parameter of the environment.
This observation constitutes an interesting feature of the quantum noises which have been seldom discussed for coupled gain-loss systems 
\cite{sc, arga, bhe, stable}.   

With the above understandings, one will find how well the phonon laser can operate. 
In analogue to an optical laser \cite{qo}, the phonon laser dynamical equations similar to those in \cite{l-4} are independently found as  
\begin{eqnarray}
\dot{b}_s&=&(-\gamma_m-i\omega_m)b_s-(1/2) ig_m p,\nonumber\\
\dot{p} &=&(1/2) ig_m \Delta N(t) b_s+\big(1/2(g-\gamma)-2 iJ\big )p,
\label{laser}
\end{eqnarray}
where $b_s=\langle \hat{b}_s\rangle$ [the subscript ``$s$" indicates the stimulated phonon mode 
to be distinguished from the thermal phonon mode in Eq. (\ref{dynamic})] and $p=\langle \hat{o}_2^\dagger\hat{o}_1\rangle$. Corresponding to the semi-classical treatment of atomic level transitions, by which the atomic levels are described quantum mechanically while the radiations are regarded as classical, we approximate the phonon laser mode in Eq. (\ref{laser}) as a mean field but insert the inversion $\Delta N(t)$ determined in a completely quantum way from Eq. (\ref{dynamic}) into the same equations. The amplification rates of the stimulated phonon field numerically found with the above equations are illustrated in Fig. 5. The threshold drive 
intensity $E_{th}$ for realizing phonon field amplification becomes lower with increased gain rate $g$, which is upper bounded in reality due to gain saturation. Under the optimal transition and optimal population inversion condition as in Figs. 2(a)-2(b), adding optical gain medium into one cavity can enhance the phonon lasing further. 

\begin{figure}[t!]
\vspace{0cm}
\centering
\epsfig{file=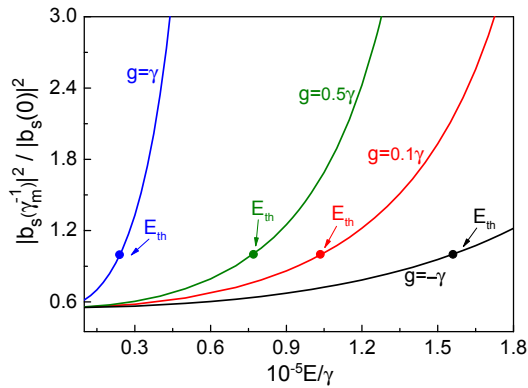,width=0.8\linewidth,clip=} 
{\vspace{-0.2cm}\caption{(color online) Amplification of the stimulated phonon field intensity in terms of the ratio between their values at $t=\gamma_m^{-1}$ and $t=0$. The system parameters are the same as those in Fig. 2(a).  }}
\vspace{0.cm}
\end{figure}

In summary, we have presented a dynamical approach to the phonon laser model of coupled active-passive resonators, which only uses a single approximation in Eq. (\ref{app}) to make the calculations of the optical supermode populations highly accurate to the system with $g_m\ll \omega_m$. Compared with a previous study based on the assumed steady states for such system \cite{jing}, we find three fundamental differences: (1) the phonon laser should operate under blue-detuned pump rather than the resonant and red-detuned ones considered in \cite{jing}---under blue-detuned drives the phonon laser performance simply betters with increased optical gain instead of reaching the optimum at the balanced gain and loss; 
(2) the phonon laser can operate even better in the $\mathcal {PT}$ symmetry broken regime ($J<(g+\gamma)/2$) in contrast to its non-existence predicted in \cite{jing}; (3) under the conditions to realize the optimum lasing, quantum noises can significantly contribute to the supermode population inversion for magnifying the stimulated phonon field. 
These features surely exist in the presence of the realistic gain saturation, though we use a model of fixed gain rate to illustrate them more clearly. According to our dynamical picture, the optimum phonon lasing in any similar setup (beyond those carrying optical gain) should be reached by choosing a proper pump detuning $\Delta$ and a suitable cavity coupling $J$, and the added optical gain highlighted in \cite{jing} will not help the laser action unless the pump detuning is within the appropriate range. For the experimentally realizable optomechanical systems ($g_m/\gamma \ll 1$), the approach can be applied to quantum dynamical processes in the blue-detuned regime, where the previously available approach of classical dynamics (see Sec. VIII in \cite{om}) is unable to deal with the problems involving quantum noises. This approach of linearizing the dynamical equations of 
weak nonlinear systems without relying on their steady states may be applied to solve other dynamical problems. 

\vspace{0cm}
 M. X. acknowledges partial funding supports from NBRPC (Grant No. 2012CB921804) and NSFC (Grant No. 61435007).
 
\begin{widetext}
\section*{Supplementary Materials}

\subsection{Approximation for weakly coupled optomechanical systems}
\renewcommand{\theequation}{S-I-\arabic{equation}}
\setcounter{equation}{0}

We start from Eq. (8) in the main text. There taking the interaction picture with respect to the Hamiltonian $H_S(t)$ is equivalent to the 
following factorization (see (2.189) in \cite{fac} or the appendices of \cite{fac1,fac2}):
\begin{eqnarray}
U(t)=\mathcal{T}\exp\{-i\int_0^t d\tau H(\tau)\}&=&\underbrace{\mathcal{T}\exp\{-i\int_0^t d\tau H_S(\tau)\}}_{U_0(t)}~\mathcal{T}\exp\{-i\int_0^t d\tau U_0^\dagger(\tau)\{H_{OM}+H_{SR}(\tau)\}U_0(\tau)\}\nonumber\\
&=&\mathcal{T}\exp\{-i\int_0^t d\tau H_S(\tau)\}~\mathcal{T}\exp \big\{-i\int_0^t d\tau \big(H_1(\tau)+H_2(\tau)\big)\big\},
\label{fact}
\end{eqnarray}
where $H_1(t)$ and $H_2(t)$ are given in Eq. (9) and Eq. (10) of the main text, respectively. The supermodes $\hat{o}_1, \hat{o}_2$ appearing in $H_1(t)$ and $H_2(t)$ are the orthogonal eigenstates of the Hermitian Hamiltonian $\omega _{c}\hat{a}_{1}^{\dagger }\hat{a}_{1}+\omega _{c}\hat{a}_{2}^{\dagger }\hat{a}_{2}+J(\hat{a}_{1}\hat{a}_{2}^{\dagger }+\hat{a}_{1}^{\dagger }\hat{a}_{2})$. Then, we take another factorization of the last time-ordered exponential 
in (\ref{fact}) as \cite{fac1,fac2}
\begin{eqnarray}
\mathcal{T}\exp \big\{-i\int_0^t d\tau \big(H_1(\tau)+H_2(\tau)\big)\big\}=\underbrace{\mathcal{T}\exp\{-i\int_0^t d\tau U_2(t,\tau)H_1(\tau)U_2^\dagger(t,\tau)\}}_{U_1(t)}~\underbrace{\mathcal{T}\exp\{-i\int_0^t d\tau H_2(\tau)\}}_{U_2(t)},
\end{eqnarray}
where $U_2(t,\tau)=\mathcal{T}\exp\{-i\int_\tau^t dt' H_2(t')\}$. Because of the non-commutativity of the effective Hamiltonian $H_1(t)$ and $H_2(t)$, 
the unitary operation $U_2(t,\tau)$ inside the first time-ordered exponential modifies the system 
mode operators in $H_1(t)$, e.g.
\begin{eqnarray}
 U_2(t,\tau)\hat{o}_1 U^\dagger_2(t,\tau)&=&\hat{o}_1+\big\{\frac{g_m}{2\omega_m}(e^{-i\omega_m t}-e^{-i\omega_m \tau})\hat{o}_1\hat{b}
-\frac{g_m}{2\omega_m}(e^{i\omega_m t}-e^{i\omega_m \tau})\hat{o}_1\hat{b}^\dagger\nonumber\\
 &+& \frac{g_m}{2(2J-\omega_m)}(e^{-i(2J-\omega_m) t}-e^{-i(2J-\omega_m) \tau})\hat{o}_2\hat{b}\nonumber\\
&+&\frac{g_m}{2(2J+\omega_m)}(e^{-i(2J+\omega_m) t}-e^{-i(2J+\omega_m) \tau})\hat{o}_2\hat{b}^\dagger\big\}+\cdots
\label{corr}
\end{eqnarray}
From the above expression up to the first order of $g_m$, the corrections of the mode operators by the unitary transformation $U_2(t,\tau)$ are seen to be negligible in the weak coupling regime $g_m\ll \gamma, \omega_m, J$. Even under the resonant condition $\omega_m=2J$, their corrections in the order of the dimensionless quantity $g_m(t-\tau)$ will take effect after a significantly long period of time $\gamma t$, given a coupling constant $g_m\sim 10^{-5}\gamma$ as in our illustrated examples. Neglecting the corrections in (\ref{corr}) will therefore not affect the results we illustrate in the main text. This only approximation of neglecting such modification from $H_2(t)$ in the calculation of the supermode populations is independent of the drive intensity $E$.
With this approximation the supermode populations can be rewritten as 
\begin{eqnarray}
\langle \hat{o}_i^\dagger\hat{o}_i(t)\rangle &=&
\mbox {Tr}_{S,R}\big (U_2^\dagger(t)U_1^\dagger(t)U_0^\dagger(t)\hat{o}_i^\dagger\hat{o}_iU_0(t)U_1(t)U_2(t) \rho_S(0)\otimes \rho_R\big)\nonumber\\
&=&\mbox {Tr}_{S,R}\big (U_1^\dagger(t)U_0^\dagger(t)\hat{o}_i^\dagger\hat{o}_iU_0(t)U_1(t) U_2(t)\rho_S(0)\otimes \rho_R U_2^\dagger(t)\big)\nonumber\\
&=&\mbox {Tr}_{S,R}\big (U_1^\dagger(t)U_0^\dagger(t)\hat{o}_i^\dagger\hat{o}_iU_0(t)U_1(t) \rho_S(0)\otimes \rho_R \big),
\end{eqnarray}
where the action $U_2(t)$ leaves the quantum state $\rho_S(0)\otimes\rho_R$ invariant because the initial state $\rho_S(0)$ is the product of a cavity vacuum state and the mechanical thermal state $\sum_{n=0}^\infty n_{th}^n/(1+n_{th})^{n+1}|n\rangle_m\langle n|$, to have $H_2(t)|0\rangle_c=0$ for the cavity vacuum state $|0\rangle_c$. The evolved supermode populations $\langle \hat{o}_i^\dagger\hat{o}_i(t)\rangle$ are only due to the successive actions of $U_0(t)$ and $U_1(t)$.

\subsection{Difference between the optical supermodes and the eigenstates of a non-Hermitian Hamiltonian}
\renewcommand{\theequation}{S-II-\arabic{equation}}
\setcounter{equation}{0}

The model of $\mathcal {PT}$ symmetric quantum mechanics with active-passive coupler is often based on the non-Hermitian Hamiltonian
\begin{eqnarray}
&&H_{PT}=(\omega_c-i\gamma) \hat{a}_1^\dagger\hat{a}_1+(\omega_c+ig) \hat{a}_2^\dagger\hat{a}_2+J(\hat{a}_1^\dagger \hat{a}_2+\hat{a}_2^\dagger\hat{a}_1)\nonumber\\
&=& (\hat{a}_1^\dagger,\hat{a}_2^\dagger)\left( 
\begin{array}{cc}
\omega_c-i\gamma & J \\ 
J & \omega_c+ig
\end{array}
\right)\left( 
\begin{array}{c}
\hat{a}_1 \\ 
\hat{a}_2%
\end{array}
\right)\nonumber\\
&=& (\hat{q}_1^\dagger,\hat{q}_2^\dagger)\left( 
\begin{array}{cc}
-\frac{i(g+\gamma)+\sqrt{4J^2-(g+\gamma)^2}}{2J} & 1 \\ 
-\frac{i(g+\gamma)-\sqrt{4J^2-(g+\gamma)^2}}{2J} & 1
\end{array}
\right)^{T,-1}\left( 
\begin{array}{cc}
\omega_c-i\gamma & J \\ 
J & \omega_c+ig
\end{array}
\right)\left( 
\begin{array}{cc}
-\frac{i(g+\gamma)+\sqrt{4J^2-(g+\gamma)^2}}{2J} & 1 \\ 
-\frac{i(g+\gamma)-\sqrt{4J^2-(g+\gamma)^2}}{2J} & 1
\end{array}
\right)^{T}\left( 
\begin{array}{c}
\hat{q}_1 \\ 
\hat{q}_2%
\end{array}
\right)\nonumber\\
&=&(\hat{q}_1^\dagger,\hat{q}_2^\dagger)\left( 
\begin{array}{cc}
\omega_c-\sqrt{J^2-(\frac{g+\gamma}{2})^2}+\frac{1}{2}(g-\gamma)i &  \\ 
 & \omega_c+\sqrt{J^2-(\frac{g+\gamma}{2})^2}+\frac{1}{2}(g-\gamma)i
\end{array}
\right)\left( 
\begin{array}{c}
\hat{q}_1 \\ 
\hat{q}_2%
\end{array}
\right)\nonumber\\
&=& \big(\omega_c-\sqrt{J^2-(\frac{g+\gamma}{2})^2}+\frac{1}{2}(g-\gamma)i\big)\hat{q}_1^\dagger\hat{q}_1+\big(\omega_c+\sqrt{J^2-(\frac{g+\gamma}{2})^2}+\frac{1}{2}(g-\gamma)i\big)\hat{q}^\dagger_2\hat{q}_2.
\end{eqnarray}
The notation ``$T,-1$" means first taking the transpose and then the inverse of the matrix. The eigenmodes of the non-Hermitian Hamiltonian $H_{PT}$ take the forms
\begin{eqnarray}
\hat{q}_1&=&-\frac{J}{\sqrt{4J^2-(g+\gamma)^2}}\hat{a}_1-\frac{i(g+\gamma)-\sqrt{4J^2-(g+\gamma)^2}}{2\sqrt{4J^2-(g+\gamma)^2}}\hat{a}_2\nonumber\\
\hat{q}_2&=&\frac{J}{\sqrt{4J^2-(g+\gamma)^2}}\hat{a}_1+\frac{i(g+\gamma)+\sqrt{4J^2-(g+\gamma)^2}}{2\sqrt{4J^2-(g+\gamma)^2}}\hat{a}_2,
\label{eigenstates}
\end{eqnarray}
from the above diagonalization procedure, and the eigenstates $\hat{q}_1^\dagger|0\rangle$ and $\hat{q}_2^\dagger|0\rangle$ are generally non-orthogonal. They can be defined as ``orthogonal" only following the definition of the inner product in $\mathcal {PT}$ symmetric quantum mechanics, i.e. $(\mathcal {PT}\hat{q}^\dagger_2|0\rangle)^T\cdot \hat{q}_1^\dagger|0\rangle=0$ where the $\mathcal {PT}$ transformation is applied to one of the vectors (see, e.g. \cite{pt}). The identification of these two eigenmodes with the orthogonal optical supermodes $\hat{o}_{1,2}=\hat{a}_1\pm \hat{a}_2$ (up to a normalization factor) is generally not true. Only for the passive setup ($g=-\gamma$) can the non-Hermitian Hamiltonian be diagonalized in terms of the orthogonal optical supermodes, i.e.
\begin{eqnarray}
H_{P}&=&(\omega_c-i\gamma) \hat{a}_1^\dagger\hat{a}_1+(\omega_c-i\gamma) \hat{a}_2^\dagger\hat{a}_2+J(\hat{a}_1^\dagger \hat{a}_2+\hat{a}_2^\dagger\hat{a}_1)\nonumber\\
&=& (\omega_c+J-i\gamma)\hat{o}_1^\dagger\hat{o}_1+(\omega_c-J-i\gamma)\hat{o}^\dagger_2\hat{o}_2.
\end{eqnarray}
The addition of the optical gain therefore leads to a significant difference. The eigenvalues of the non-Hermitian Hamiltonian $H_{PT}$ manifest in light transportation inside coupled active-passive systems. Using the differential equations 
\begin{eqnarray}
&&\frac{d}{dt}a_1=-\gamma a_1-iJ a_2 +E e^{-i\Delta t},\\
&&\frac{d}{dt}a_2=g a_2-iJ a_1
\end{eqnarray}
from taking the averages in Eqs. (1)-(2) of the main text (with $g_m=0$), one will find two transmission resonances (two values of $\Delta$ to have peaked $|a_2(t)|^2$) when $J>(g+\gamma)/2$ and one resonance when $J<(g+\gamma)/2$ from its solution
\begin{eqnarray}
\left( 
\begin{array}{c}
a_1(t)\\ 
a_2(t)
\end{array}
\right)=\int_0^t d\tau\exp\{\left( 
\begin{array}{cc}
-\gamma & -iJ \\ 
-iJ & g
\end{array}
\right)(t-\tau)\}\left( 
\begin{array}{c}
E e^{-i\Delta \tau}\\ 
0
\end{array}
\right).
\end{eqnarray}
These resonances have been observed by experiments \cite{ex6, ex7}, and reflect the real parts of the eigenvalues of the non-Hermitian Hamiltonian $H_{PT}$. 

When it comes to the phonon induced transition between coupled cavity modes in the presence of the optomechanical coupling $g_m\neq 0$, one should clarify whether such transition could take place between the states $\hat{q}_1^\dagger|0\rangle$ and $\hat{q}_2^\dagger|0\rangle$. Since they are non-orthogonal with an overlap $\langle 0|\hat{q}_1\hat{q}_2^\dagger|0\rangle\neq 0$, the transition between these two eigenmodes as the superpositions of the cavity modes $\hat{a}_1$ and $\hat{a}_2$ is impossible. Otherwise a trivial Hamiltonian in the form of identity operator can cause the automatic transition between them, being contradictory with the available observations. 
In this sense the action of the Hermitian Hamiltonian $H_{OM}=-g_m\hat{a}_1^\dagger\hat{a}_1(\hat{b}+\hat{b}^\dagger)$ of optomechanical coupling can not lead to the transition between two non-orthogonal supermode states. In phonon lasing the stimulated phonon field is amplified due to the effective transition between the orthogonal states $\hat{o}_1^\dagger|0\rangle$ and $\hat{o}_2^\dagger|0\rangle$ under the action of $H_{OM}$. 

\subsection{Calculation of the optical supermode populations}
\renewcommand{\theequation}{S-III-\arabic{equation}}
\setcounter{equation}{0}

The main equations, Eq. (14) in the main text, for the determination of the supermode populations can be found following Eq. (11.2.33) in \cite{book} (in the absence of the amplification part) or Eq. (A4) in \cite{bhe}. Their extended forms including the differential equations for the conjugated operators read
\begin{eqnarray}
\frac{d}{dt}\hat{o}_1&=& \frac{1}{2}(g-\gamma)\hat{o}_1+\frac{1}{2}(g+\gamma) e^{2iJt}\hat{o}_2+ ig_m E_1(t)e^{iJt}(\hat{b}%
e^{-i\omega_m t}+\hat{b}^\dagger e^{i\omega_m t}) \nonumber\\
&+&\underbrace{\big(\gamma E_1(t)-g E_2(t)\big)e^{iJt}}_{\lambda_1(t)}+\underbrace{\sqrt{g}
e^{iJt}e^{i\omega_c t}\hat{\xi}^\dagger_a(t)+\sqrt{\gamma}
e^{iJt}e^{i\omega_c t}\hat{\xi}_p(t)}_{\hat{n}_1(t)},  \nonumber \\
\frac{d}{dt}\hat{o}^\dagger_1&=& \frac{1}{2}(g-\gamma)\hat{o}^\dagger_1+\frac{1}{2}(g+\gamma) e^{-2iJt}\hat{o}^\dagger_2- ig_m
E^\ast_1(t)e^{-iJt}(\hat{b}e^{-i\omega_m t}+\hat{b}^\dagger e^{i\omega_m t})\nonumber\\
&+& \big(\gamma E^\ast_1(t)-g E^\ast_2(t)\big)e^{-iJt}+\sqrt{g}e^{-iJt}e^{-i\omega_c t}\hat{\xi}_a(t)+
\sqrt{\gamma}e^{-iJt}e^{-i\omega_c t}\hat{\xi}^\dagger_p(t), \nonumber \\
\frac{d}{dt}\hat{o}_2&=& \frac{1}{2}(g+\gamma)e^{-2iJt}\hat{o}_1+\frac{1}{2}(g-\gamma)\hat{o}_2+ig_m E_1(t)e^{-iJt}(\hat{b}e^{-i\omega_m t}+\hat{b}^\dagger e^{i\omega_m t})\nonumber\\
&+& \underbrace{\big(\gamma E_1(t)+g E_2(t)\big)e^{-iJt}}_{\lambda_2(t)}+\underbrace{\sqrt{g} e^{-iJt}e^{i\omega_c t}\hat{\xi}^\dagger_a(t)-\sqrt{\gamma}e^{-iJt}e^{i\omega_c t}\hat{\xi}_p(t)}_{\hat{n}_2(t)}, \nonumber \\
\frac{d}{dt}\hat{o}^\dagger_2&=& \frac{1}{2}(g+\gamma)e^{2iJt}\hat{o}^\dagger_1+\frac{1}{2}(g-\gamma)\hat{o}^\dagger_2-ig_m
E^\ast_1(t)e^{iJt}(\hat{b}e^{-i\omega_m t}+\hat{b}^\dagger e^{i\omega_m t})\nonumber\\
&+& \big(\gamma E^\ast_1(t)+g E^\ast_2(t)\big)e^{iJt}+
\sqrt{g} e^{iJt}e^{-i\omega_c t}\hat{\xi}_a(t)-\sqrt{\gamma}e^{iJt}e^{-i\omega_c t}\hat{\xi}^\dagger_p(t), 
\nonumber \\
\frac{d}{dt}\hat{b}&=&-\gamma_m \hat{b}+ig_m E^\ast_1(t)e^{i\omega_m t}(\hat{%
o}_1 e^{-iJt}+\hat{o}_2 e^{iJt})+ ig_m E_1(t)e^{i\omega_m t}(\hat{o}%
^\dagger_1 e^{iJt}+\hat{o}^\dagger_2 e^{-iJt})  \nonumber \\
&+&\underbrace{\sqrt{2\gamma_m}e^{i\omega_m t}\hat{\xi}_m(t)}_{\hat{n}_3(t)}+\underbrace{2ig_m|E_1(t)|^2e^{i\omega_m t}}_{\lambda_3(t)},
\nonumber \\
\frac{d}{dt}\hat{b}^\dagger&=&-\gamma_m \hat{b}^\dagger -ig_m
E^\ast_1(t)e^{-i\omega_m t}(\hat{o}_1 e^{-iJt}+\hat{o}_2 e^{iJt})-ig_m
E_1(t)e^{-i\omega_m t}(\hat{o}^\dagger_1 e^{iJt}+\hat{o}^\dagger_2 e^{-iJt})
\nonumber \\
&+&\sqrt{2\gamma_m}e^{-i\omega_m t}\hat{\xi}^\dagger_m(t)-2ig_m|E_1(t)|^2e^{-i\omega_m t},  \label{main}
\end{eqnarray}
where $E_1(t), E_2(t)$ are given in Eq. (7) of the main text.
We write the above equations in terms of a $6\times 6$ dynamical matrix $\hat{M}(t)$ as 
\begin{eqnarray}
\frac{d}{dt}\hat{\vec{c}}(t)=\hat{M}(t)\hat{\vec{c}}(t)+\vec{\lambda}(t)+\hat{\vec{n}}(t),
\label{equations}
\end{eqnarray}
where 
\begin{eqnarray}
\hat{\vec{c}}(t)&=&(\hat{o}_1(t),\hat{o}_1^\dagger(t),\hat{o}_2(t),\hat{%
o}_2^\dagger(t),\hat{b}(t),\hat{b}^\dagger(t))^T,\nonumber\\
\vec{\lambda}(t)&=& ( \lambda_1(t), \lambda^\ast_1(t),\lambda_2(t), \lambda^\ast_2(t),\lambda_3(t),\lambda^\ast_3(t))^T\nonumber\\
\hat{\vec{n}}(t)&=&(\hat{n}_1(t),\hat{n}_1^\dagger(t),\hat{n}_2(t),\hat{n}_2^\dagger(t),\hat{n}%
_3(t),\hat{n}_3^\dagger(t))^T. 
\end{eqnarray}
The enhancement of beam-splitter type coupling or squeezing type coupling, as we describe in the main text, can be realized by adjusting the elements of the dynamical matrix $\hat{M}(t)$. 

The general solution of the above dynamical equations is 
\begin{eqnarray}
\hat{\vec{c}}(t)&=&\mathcal{T}\exp \{\int_0^t d\tau \hat{M}(\tau)\}\hat{\vec{c}}%
(0)+\int_0^t d\tau \mathcal{T}\exp \{\int_\tau^t dt^{\prime} \hat{M}(t^{\prime})\}\big(\vec{\lambda}(\tau)+\hat{\vec{n}}(\tau)\big)= \hat{\vec{c}}_s(t)+\vec{c}_{ds}(t)+\hat{\vec{c}}_n(t).
\label{sol}
\end{eqnarray}
The operator $\mathcal{T}\exp \{\int_0^t d\tau \hat{M}(\tau)\}$ is numerically calculated as the product $\prod_{i=N-1}^0 (I+\hat{M}(\tau_i)h)$, where the range $[0,t]$ is divided into $N$ pieces with the step size $h$. The step size $h$ is chosen to be sufficiently small so that the matrix product becomes insensitive to it. The time ordered exponential in the general form $\mathcal{T}\exp\{\int_\tau^t dt' \hat{M}(t^{\prime})\}$ can be found with such matrix products and their inverses, and is represented as a $6\times 6$ matrix
\begin{eqnarray}
\mathcal{T}\exp\{\int_\tau^t dt' \hat{M}(t^{\prime})\}= \left( 
\begin{array}{cccccc}
d_{11}(t,\tau) & d_{12}(t,\tau) & \cdots &  &  & d_{16}(t,\tau) \\ 
d_{21}(t,\tau) & d_{22}(t,\tau) & \cdots &  &  & d_{26}(t,\tau) \\ 
\vdots & \vdots & \ddots &  &  & \vdots  \\ 
d_{61}(t,\tau) & d_{62}(t,\tau) & \cdots &  &  & d_{66}(t,\tau) \\ 
&  &  &  &  & 
\end{array}
\right).
\label{t-exp}
\end{eqnarray}

There are three terms in the solution (\ref{sol}) of Eq. (\ref{main}). The supermode populations from the first term are obtained by taking the average of $\hat{o}_{i,s}^\dagger\hat{o}_{i,s}(t)$ with respect to the system's initial state $|0\rangle_c\langle 0|\otimes \sum_n\frac{n_{th}^n}{(1+n_{th})^{n+1}}|n\rangle_m\langle n|$, the product of the cavity vacuum state and the mechanical thermal state, where $n_{th}$ is the thermal reservoir mean occupation number. This part of the contribution is found as
\begin{eqnarray}
\langle\hat{o}^\dagger_{1,s}\hat{o}_{1,s}(t)\rangle=
d_{21}(t,0)d_{12}(t,0)+d_{23}(t,0)d_{14}(t,0)+d_{25}(t,0)d_{16}(t,0)(n_{th}+1)+d_{26}(t,0)d_{15}(t,0)n_{th},
\label{s1}
\end{eqnarray}
\begin{eqnarray}
\langle\hat{o}^\dagger_{2,s}\hat{o}_{2,s}(t)\rangle=
d_{41}(t,0)d_{32}(t,0)+d_{43}(t,0)d_{34}(t,0)+d_{45}(t,0)d_{36}(t,0)(n_{th}+1)+d_{46}(t,0)d_{35}(t,0)n_{th}.
\label{s2}
\end{eqnarray}
Together with the displacement terms due to the action of $U_0(t)$, the second pure drive term of $\vec{\lambda}(t)$ gives rise to 
the following contribution
\begin{eqnarray}
\langle\hat{o}^\dagger_{1,ds}\hat{o}_{1,ds}(t)\rangle =\big|E_1(t)+E_2(t)+o_{1,ds}(t)\big|^2,
~~~~\langle\hat{o}^\dagger_{2,ds}\hat{o}_{2,ds}(t)\rangle =\big|E_1(t)-E_2(t)+o_{2,ds}(t)\big|^2,
\end{eqnarray}
where
\begin{eqnarray}
o_{1,ds}(t) &=&\int_0^t d\tau   \big\{ d_{11}(t,\tau)\big(\gamma E_1(\tau)-g E_2(\tau)\big)e^{iJ\tau}+d_{12}(t,\tau)\big(\gamma E^\ast_1(\tau)-g E^\ast_2(\tau)\big)e^{-iJ\tau}\nonumber\\
&+&d_{13}(t,\tau)\big(\gamma E_1(\tau)+g E_2(\tau)\big)e^{-iJ\tau}+d_{14}(t,\tau)\big(\gamma E^\ast_1(\tau)+g E^\ast_2(\tau)\big)e^{iJ\tau}\nonumber\\
&+&2ig_m d_{15}(t,\tau)|E_2(\tau)|^2e^{i\omega_m \tau}-2ig_md_{16}(t,\tau)|E_2(\tau)|^2e^{-i\omega_m \tau}\big\},
\end{eqnarray}
\begin{eqnarray}
o_{2,ds}(t)&= &\int_0^t d\tau   \big\{ d_{31}(t,\tau)\big(\gamma E_1(\tau)-g E_2(\tau)\big)e^{iJ\tau}+d_{32}(t,\tau)\big(\gamma E^\ast_1(t)-gE^\ast_2(\tau)\big)e^{-iJ\tau}\nonumber\\
&+&d_{33}(t,\tau)\big(\gamma E_1(\tau)+g E_2(\tau)\big)e^{-iJ\tau}+d_{34}(t,\tau)\big(\gamma E^\ast_1(\tau)+g E^\ast_2(\tau)\big)e^{iJ\tau}\nonumber\\
&+&2ig_m d_{35}(t,\tau)|E_2(\tau)|^2e^{i\omega_m \tau}-2ig_md_{36}(t,\tau)|E_2(\tau)|^2e^{-i\omega_m \tau}\big\}.
\end{eqnarray}
Here the contributions from the drive terms $\vec{\lambda}(t)$ to the cavity modes add to those of $E_1(t), E_2(t)$, which are obtained without amplification and dissipation, to have the correct ones under the decoherence effects. The effect of the drive terms is more obvious when comparing the results of with and without decoherence in the situation of $g_m=0$. Finally, the contribution from the noise drive terms $\hat{\vec{n}}(t)$ can be found with the relations $\langle \hat{\xi}_i(t) \hat{\xi}_i^\dagger (t')\rangle=\delta(t-t')$, $\langle \hat{\xi}^\dagger_i(t) \hat{\xi}_i (t')\rangle=0$ for $i=p,a$ and $\langle \hat{\xi}_m(t) \hat{\xi}_m^\dagger (t')\rangle=(n_{th}+1)\delta(t-t')$, $\langle \hat{\xi}^\dagger_m(t) \hat{\xi}_m (t')\rangle=n_{th}\delta(t-t')$ as the expectation values over the reservoir states, to have
\begin{eqnarray}
\langle \hat{o}^\dagger_{1,n}(t)\hat{o}_{1,n}(t)\rangle&=&
 \gamma\int_0^t d\tau d_{21}(t,\tau)d_{12}(t,\tau)+g\int_0^t d\tau
d_{22}(t,\tau)d_{11}(t,\tau)+\gamma\int_0^t d\tau
d_{23}(t,\tau)d_{14}(t,\tau)  \nonumber \\
&+&g\int_0^t d\tau d_{24}(t,\tau)d_{13}(t,\tau)+2\gamma_m
(n_{th}+1)\int_0^t d\tau d_{25}(t,\tau)d_{16}(t,\tau)  \nonumber \\
&+&2\gamma_m n_{th}\int_0^t d\tau d_{26}(t,\tau)d_{15}(t,\tau),
\label{n1}
\end{eqnarray}
\begin{eqnarray}
\langle \hat{o}^\dagger_{2,n}(t)\hat{o}_{2,n}(t)\rangle&=& \gamma\int_0^t d\tau d_{41}(t,\tau)d_{32}(t,\tau)+g\int_0^t d\tau
d_{42}(t,\tau)d_{31}(t,\tau)+\gamma\int_0^t d\tau
d_{43}(t,\tau)d_{34}(t,\tau)  \nonumber \\
&+&g\int_0^t d\tau  d_{44}(t,\tau)d_{33}(t,\tau)+2\gamma_m
(n_{th}+1)\int_0^t d\tau d_{45}(t,\tau)d_{36}(t,\tau)  \nonumber \\
&+&2\gamma_m n_{th}\int_0^t d\tau d_{46}(t,\tau)d_{35}(t,\tau).
\label{n2}
\end{eqnarray}

Adding the three parts of contributions together gives the total supermode populations. The detailed contributions from the different parametric processes or different noises can be identified by the evolutions of the matrix elements $d_{ij}(t,0)$, which also specify the temporal distributions of the general matrix elements $d_{ij}(t,\tau)$. For a passive setup without optical gain, the amplification noise contributions proportional to $g$ in Eqs. (\ref{n1}) and (\ref{n2}) will be replaced by the ones from another part of dissipation noise contributions. Under the conditions as in Figs. 2(a) and 2(b) of the main text, the term with the factor $(n_{th}+1)$ in Eq. (\ref{n1}) is found to contribute to the blue supermode population significantly.

\end{widetext}

\end{document}